\documentclass[twocolumn,superscriptaddress,floatfix,preprintnumbers,amssymb
,amsmath,showpacs]{revtex4}
\usepackage{graphicx}
\usepackage{dcolumn}
\usepackage{bm}
\usepackage[latin1]{inputenc}
\usepackage[mathscr]{eucal}
\usepackage{epsfig}
\usepackage{epsf}

\begin{document}

\title{Non-adiabadic charge pumping in a hybrid SET transistor}

\author{Dmitri V.~Averin}
\affiliation{Department of Physics and Astronomy, Stony Brook
University, SUNY, Stony Brook, NY 11794-3800 }

\author{Jukka P. Pekola}
\affiliation{Low Temperature Laboratory, Helsinki University of
Technology, P.O. Box 3500, 02015 TKK, Finland}

\begin{abstract}
We study theoretically current quantization in the charge turnstile
based on the hybrid (SINIS or NISIN) SET transistor. The
quantization accuracy is limited by either Andreev reflection or by
Cooper pair - electron cotunneling. The rates of these processes are
calculated in the ``above-the-threshold'' regime when they compete
directly with the lowest-order tunneling. We show that by shaping
the ac gate voltage driving the turnstile, it should be possible to
achieve the metrological accuracy of $10^{-8}$, while maintaining
the absolute value of the quantized current on the order of 30 pA,
just by one turnstile with realistic parameters using aluminium as
superconductor.
\end{abstract}


\maketitle

\begin{figure}
\setlength{\unitlength}{1.0in}
\begin{picture}(3.3,.8)
\put(.0,-.2){\epsfxsize=3.3in \epsfbox{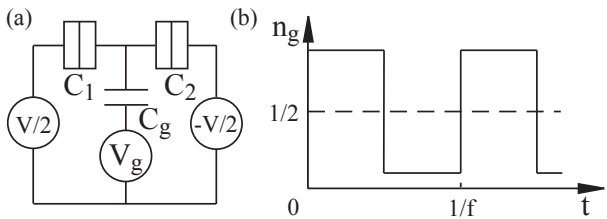}}
\end{picture}
\caption{(a) Hybrid SET transistor with SIN or NIS tunnel junctions,
and (b) time dependence of the ac gate-induced charge $n_g= C_g V_g
/e$ oscillating with frequency $f$ around the point $n_g=1/2$.}
\label{f0}
\end{figure}

Nanoscale tunneling structures provide the general basis for
development of metrological sources of electrical current utilizing
controlled transfer of individual charges \cite{al91}. However,
despite the beautiful achievements based on experiments with gated
arrays of metallic tunnel junctions \cite{b2,b3,kel99,b10}, and with
semiconductor surface-acoustic-wave and charge-coupled devices
\cite{sh96,fuj04,bl07,ka07}, no fully satisfactory system in terms
of both the accuracy and current magnitude has been realized yet. It
was suggested recently \cite{b1} that an unexpectedly simple
structure, a single-electron (SET) transistor with two hybrid normal
metal - superconductor (NIS) or superconductor - normal metal (SIN)
tunnel junctions holds promise as a quantized source of current. The
first experiments with such a transistor as a turnstile \cite{b1}
demonstrated correct operation at the level of classical charge
dynamics, but they were not yet conclusive as to its ultimate
accuracy. In this Letter we analyze theoretically all the relevant
higher-order quantum tunneling processes which limit this accuracy.
The main conclusion we reach is that these errors can be suppressed
in a single ordinary aluminium-based device to the level mandated by
the metrological requirements ($\leq 10^{-8}$), while keeping the
absolute current relatively large (see Fig.~\ref{fig:comb} below),
provided the single-electron charging energy of the turnstile is
sufficiently high.

\begin{figure}
\setlength{\unitlength}{1.0in}
\begin{picture}(3.2,1.3)
\put(.4,-.1){\epsfxsize=2.5in \epsfbox{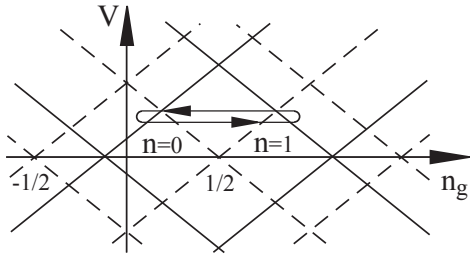}}
\end{picture}
\caption{Charge stability diagram of the hybrid SET transistor
operated as a turnstile. Dashed lines are the tunneling thresholds
of the rhombic stability regions $n=0,1$ in the normal-metal case.
Solid lines show the thresholds in the hybrid transistor shifted by
the superconducting gap $\Delta$. Periodic variation of the gate
voltage (line with arrows) transfers one electron per period through
the transistor.} \label{f1}
\end{figure}

The basic ``classical'' dynamics of the hybrid SET transistor
(Fig.~\ref{f0}) as a charge turnstile can be described conveniently
on the stability diagram shown in Fig.~\ref{f1}. Periodic variation
of the gate-induced charge $n_g (t) \equiv C_g V_g(t)/e$ with
frequency $f$ (notations are defined by Fig.~\ref{f0}) indicated by
the line with arrows in Fig.~\ref{f1} drives the transistor
periodically between the two nearest stability areas, e.g., $n=0$
and $n=1$, where $n$ is the equilibrium number of extra electrons on
the island. The turnstile operation requires that the lowest-order
tunneling transitions are organized so that at finite bias voltage
$V$ and low temperature $T$ they transfer precisely one electron per
period $1/f$ through the transistor \cite{b1}. The properties of the
tunneling thresholds (solid lines in Fig.~\ref{f1}) that make this
possible in the hybrid transistor but not in the normal-metal one
can be seen from Fig.~\ref{f1}. The thresholds in the hybrid are
shifted with respect to the normal-metal system (dashed lines in
Fig.~\ref{f1}) by the superconducting energy gap $\Delta$, i.e., the
shift along the $n_g$ axis is $\delta=\Delta/2E_C$, where $E_C
\equiv e^2/2C_{\Sigma}$ and $C_{\Sigma}=C_1+C_2+C_g$, expanding the
stability areas. As a result, the neighboring stability areas
overlap, and the gate voltage can drive the system between them
keeping it all the time in the region of suppressed tunneling. Also,
in this process, when the outgoing gate voltage trajectory leaves
the initial stability area, it crosses only one of the tunneling
thresholds that define this area, allowing electron tunneling in
only one direction. For instance, if the state $n=0$ is brought by
increase of $n_g$ out of its expanded stability area into the $n=1$
area (Fig.~\ref{f1}), electron can tunnel into the transistor island
only through the left junction. When the gate voltage decreases back
to $n=0$ state, electron can tunnel out only through the right
junction \cite{b1}.

This turnstile operation is possible for any, e.g. harmonic,
periodic time dependence $n_g(t)$ with the amplitude sufficiently
large to move the system between the two stability areas
(Fig.~\ref{f1}). The time that the system spends, however, in the
overlap region of the two areas does not play any useful role in the
turnstile dynamics, and on the contrary, increases the effect of the
unwanted transitions. In order to maximize the turnstile operation
frequency and the output current, one needs then to minimize this
time by making the waveform $n_g(t)$ as in Fig.~\ref{f0}b. In this
case, the system is switched abruptly between the regions where
electron tunnels in or out of the transistor, and the operation
frequency $f$ is limited only by the need to make the probability of
missing these transitions $e^{-\gamma /2f}$ sufficiently small. At
zero temperature, the corresponding tunneling rate is $\gamma (U)
=\gamma_0 (U^2/\Delta^2-1)^{1/2}$, where $U$ is electrostatic energy
change due to tunneling, $\gamma_0 \equiv G \Delta/e^2$ and $G$ is
the junction tunnel conductance. Optimized waveform (Fig.~\ref{f0}b)
should be abrupt on the time scale of the turnstile period $1/f$. It
should, however, be smooth on the scale $h/\Delta$ to avoid
excitations of the higher-energy states of the transistor leading to
errors in the turnstile dynamics. This condition can be satisfied
easily, since for a typical current of 100 pA, the frequency $f=I/e
< 1$ GHz is well below $\Delta/h \simeq 50$ GHz.

In addition to missed cycles of tunneling due to finite frequency
$f$, the basic correct tunneling sequence can be interrupted by
thermal excitations due to finite temperature $T$, or quantum
higher-order tunneling processes \cite{b4} which set the theoretical
limit on the accuracy of the quantized current $I=ef$ produced by
the turnstile. The rate of thermal errors depends on how far the
gate-voltage trajectory is from the crossing points of the four
relevant tunneling thresholds shown as solid lines in Fig.~\ref{f1}.
The thresholds are given by the conditions $U_j^{\pm}=\Delta$ on
electrostatic energy change $U_j^{\pm}$ due to forward (wanted) or
backward (unwanted) electron tunneling in the $j$th junction:
\[ U_1^{\pm}=\pm 2E_C(v_1+n_g-1/2) , \;\; U_2^{\pm}=\pm
2E_C(v_2-n_g+1/2)\, , \] where $v_j$ is a part of $V$ that drops
across the $j$th junction: $v_1= (C_2+C_g/2)V/e$ and $v_2=
(C_1+C_g/2)V/e$. These equations show that at the thresholds of
correct tunneling, the energy barriers for unwanted transitions
through the opposite junction of the transistor are
$\Delta-U_1^-=\Delta-U_2^-=eV$. Thus, with exponential accuracy, the
thermal probability of electron tunneling in or out through the
wrong junction leading to no net charge transfer in the cycle, is
$e^{-eV/k_BT}$. Another type of unwanted thermal transitions is the
excitation of an extra electron through the transistor during the
part of the period spent in the overlap region of the two stability
areas. Electron is transferred by two successive excitations over
the energy barriers $\Delta-U_j^+$, so that the thermal excitation
exponent for the overall process is $e^{-(2\Delta-eV)/k_BT}$.
Comparing the probabilities of the two types of thermal errors, we
see that the thermal error rate is minimum for $eV \simeq \Delta$:
in practise, the resulting classical error $e^{-\Delta/k_BT}$ is
less than $10^{-8}$ at realistic temperatures $T\simeq 100$ mK.

\begin{figure}
\setlength{\unitlength}{1.0in}
\begin{picture}(3.2,1.3)
\put(.4,-.1){\epsfxsize=2.5in \epsfbox{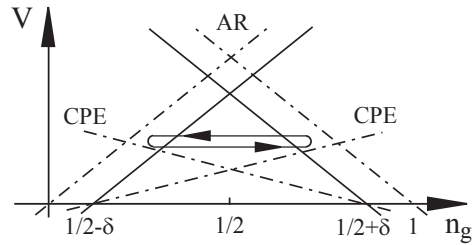}}
\end{picture}
\caption{Tunneling thresholds for several different tunneling
processes: Andreev reflection (AR) and Cooper-pair/electron (CPE)
cotunneling, in a hybrid SET transistor. As in \protect
Fig.~\ref{f1}, solid lines are the thresholds of the lowest-order
tunneling. All tunneling processes are driven by bias voltage $V$
and are allowed above the corresponding threshold.} \label{f2}
\end{figure}

We consider now quantum errors assuming ideal s-wave BCS
superconductors in the hybrid transistor structure. The rates of
``elastic'' higher-order processes which transfer electrons
coherently, without creating excitations in the electrodes, are
different in the NISIN and SINIS structures. In the NISIN
transistor, the dominant elastic process is electron cotunneling,
the rate of which is smaller than the rate $\gamma$ of the
lowest-order tunneling $\gamma$ roughly by a factor $(\hbar
G/e^2)(\delta E/\Delta)$ \cite{b5}, where $\delta E/\Delta$ is the
level spacing of the transistor island. For typical parameters,
e.g., $\mu$m-size island, this suppression factor is very small,
about $10^{-6} - 10^{-7}$, but does not quite reach the
metrologically required level. In the SINIS transistor, in the
relevant regime $eV \simeq \Delta$, the main contribution to elastic
leakage is due to rectification of the ac Josephson current through
the transistor. The resulting dc current is proportional to the
square of the SINIS critical current and is much smaller than the
inelastic leakage assisted by Andreev reflection that is considered
below.

The rates of incoherent ``inelastic'' processes depend only on the
local properties of the tunnel junctions and are the same in the
NISIN and SINIS transistors. Intensity of these processes decreases
rapidly with the number of involved electron transfers. The simplest
process of electron inelastic cotunneling through the transistor is
energetically forbidden in the relevant voltage range $eV< 2\Delta$.
Transitions next in the order of complexity are Andreev reflection
(AR), i.e. tunneling of two electrons in a Cooper-pair for which the
superconducting gap does not provide an energy barrier, and
Cooper-pair/electron (CPE) cotunneling. Electrostatic energy gains
in these processes are
\[ \mbox{AR:} \;\;\; U_1^{++}=4E_C(v_1+n_g-1) , \;\; U_2^{++}=
4E_C(v_2-n_g)\, , \]

\vspace{-5ex}

\begin{equation}
\mbox{CPE:} \;\;\;\; W_j^+=U_j^+ +eV,
\end{equation}
and the diagram of the corresponding thresholds, $U=0$ for AR, and
$W=\Delta$ for CPE is shown in Fig.~\ref{f2}. If single-electron
charging energy is small, $E_C<\Delta$, (i.e. $\delta >1/2$ in
Fig.~\ref{f2}) AR is allowed in the regions of the lowest-order
tunneling needed for turnstile operation. Each AR process causes an
error by transferring one uncontrolled extra electron. For larger
charging energy, $E_C>\Delta$, turnstile can be operated in the
regime with suppressed AR ($\delta <1/2$ in Fig.~\ref{f2}), and only
the higher-order CPE processes cause errors. Qualitatively, in the
CPE, instead of one electron jumping in or out of the transistor
island, this transition is combined coherently with electron
transfer of another electron through the whole transistor. To avoid
creating superconducting excitations, the necessary tunneling of two
electrons in one of the transistor junctions in this process happens
as AR. This CPE cotunneling is allowed energetically for any
turnstile parameters and limits the accuracy of current
quantization.

Quantitatively, we calculate the rates of the two higher-order
tunneling processes assuming the simple quasi-1D ballistic geometry
of the turnstile junctions, in which different transport modes in
the electrodes are not mixed by tunneling. This assumption is
reasonable in view of large conductivity of electrodes of practical
SET transistors. Because of the non-adiabatic variation of the gate
voltage (Fig.~\ref{f0}b), both higher-order tunneling processes take
place in the ``above-the-threshold'' regime, when they coexist with
the lowest-order single-particle tunneling. We start with the rate
$\gamma_{AR}$ of the {\em Andreev reflection}. Above the
single-particle threshold, the standard description of AR as the
two-step transition perturbative in the electron tunneling
amplitudes $t$ (see, e.g., \cite{wil}) should be modified to account
for the competing single-particle tunneling with rate $\gamma(U^+)$.
Similarly to the theory of the Coulomb-blockade threshold
\cite{cbt}, this can be done simply by taking into account the
lifetime broadening $i\gamma(U^+)/2$ of the initial state.

Because of the mutual coherency of Cooper pairs in different orbital
states in the superconducing electrode, the amplitudes of the
Cooper-pair tunneling from different states $p$ within each
transport mode into the two single-particle states with energies
$\epsilon_k$, $\epsilon_l$ in the normal electrode should be summed
coherently. The total AR amplitude $A$ is then:
\[ A (\epsilon_k,\epsilon_l) = \sum_p u_p v_p t_{pk}t_{pl}
(\frac{1}{\Omega_p +\epsilon_k-u}+\frac{1}{\Omega_p +\epsilon_l-u}),
\]

\vspace*{-3ex}

\begin{equation}
u=U^+ +i \gamma(U^+)/2 \, , \label{e5}
\end{equation}
where $u_p,v_p=[(1 \pm \epsilon_p/\Omega_p)/2]^{1/2}$ are the usual
BCS quasiparticle factors and $\Omega_p= (\Delta^2+ \epsilon_p^2 )^
{1/2}$ is the quasiparticle energy. Taking the sum over $p$ under
the standard approximation of constant density of states $\rho$ and
tunnel amplitudes $t$ in the relevant energy range on the order of
energy gap $\Delta$, we get
\[ A (\epsilon_k, \epsilon_l)=  \rho t^2 \Delta
[a(u-\epsilon_k)+a(u-\epsilon_l)] ,  \]

\vspace*{-3ex}

\begin{equation}
a(\epsilon)=(\epsilon^2-\Delta^2)^{-1/2}  \ln \left[
\frac{\Delta-\epsilon + (\epsilon^2-\Delta^2)^{1/2}}{\Delta-\epsilon
- (\epsilon^2-\Delta^2)^{1/2}} \right]. \label{e6}
\end{equation}
The main qualitative feature of the amplitude $A$ is the resonance
at the gap edge, $\epsilon \simeq  \Delta$, where the rate $|A|^2$
diverges as $1/|\epsilon -\Delta|$. Level broadening, in our case
due to the single-particle tunneling with rate $\gamma$, broadens
the resonance and suppresses the divergence.

The amplitude $A$ gives the total rate of AR at small temperatures
$k_B T\ll \Delta$ :
\[ \gamma_{AR}= \frac{2\pi}{\hbar}  \sum_{k,l} |A |^2
(1-f(\epsilon_k))(1-f(\epsilon_l))\delta
(\epsilon_k+\epsilon_l-U^{++})\, , \] where in the adopted quasi-1D
model the states $k,l$ in the sum should belong to the same
transport mode. The result of summation over these modes can be
expressed in terms of the normal-state conductance $G$ within the
natural junction model in which transparency $t^2$ varies
exponentially with energy on the scale $\epsilon_0 \gg \Delta$. The
effective number $\mathcal{N}$ of the transport modes in the
junction is determined then by the decrease of transparency with
increasing transverse energy of the mode: $\mathcal{N}= Sm \epsilon
_0/\pi \hbar^2$, where $S$ is the junction area and $m$ is electron
mass. The sum over modes and integration over the total energy can
then be done separately giving the AR rate:
\[  \gamma_{AR}=\frac{\gamma_0  g \Delta}{16\pi
\mathcal{N}} \int d\epsilon f(\epsilon -U^{++}/2) f(-\epsilon
-U^{++}/2) \]

\vspace*{-3ex}

\begin{equation}
\times  | \sum_{\pm} a(\pm \epsilon+E_C-i\gamma /2 ) |^2, \;\;\;
g\equiv \hbar G/e^2.  \label{e7}
\end{equation}

If AR transitions are not energetically allowed, the leakage current
is determined by the third-order CPE cotunneling which combines AR
with one more electron transfer in the opposite junction. The part
of the CPE amplitude $\mathcal{A}$  that corresponds to the
two-electron AR transfer process is calculated as above for direct
AR. Combining terms with different ordering of the three involved
electron transfers we get the total CPE amplitude
\[ \mathcal{A}= (\frac{1}{2E_C+2U^+ -\epsilon_k-\epsilon_l}
+\frac{1}{\epsilon_k+\epsilon_l-U^+ -u} ) \cdot \]

\vspace*{-4ex}

\begin{equation}
[a(\epsilon_k-U^+)+a(\epsilon_l-U^+)] + [a(u -\epsilon_k)+
 a(u-\epsilon_l)] \cdot \label{e12}
\end{equation}

\vspace*{-4ex}

\[ (\frac{1}{U^+ +u-\epsilon_k-\epsilon_l} +
\frac{1}{2E_C-U^+ -u +\epsilon_k+\epsilon_l} ) . \]
Summing all transitions with this amplitude as above, we obtain the
total rate of the CPE cotunneling:
\[ \gamma_{CPE}=\frac{\gamma_0  g^2 \Delta}{32\pi^2 \mathcal{N}}
\int_\Delta^\infty d\Omega
\frac{\Omega}{\sqrt{\Omega^2-\Delta^2}}\int d\epsilon_k \int
d\epsilon_l |\mathcal{A}|^2 \cdot \]

\vspace*{-3ex}

\begin{equation} \label{e11}
[1-f(\epsilon_k)][1-f(\epsilon_l)]f(\Omega+\epsilon_k+\epsilon_l-U^+
-eV).
\end{equation}

Figure \ref{fig:comb}a shows the gate dependence of the
zero-temperature normalized rates of the (wanted) single-particle
tunneling, $\tilde{\gamma}\equiv\gamma/\gamma_0$, of the AR
transitions, $\tilde{\gamma}_{AR}\equiv \gamma_{AR}/(\gamma_0
g/16\pi \mathcal{N})$, and CPE cotunneling,
$\tilde{\gamma}_{CPE}\equiv\gamma_{CPE}/(\gamma_0 g^2 /32\pi^2
\mathcal{N})$ at the optimum bias point $eV=\Delta$ for a few values
of the ratio $E_C/\Delta$. As in Fig.~\ref{f2}, the thresholds of
single-particle and Andreev processes coincide for $E_C/\Delta=1$,
but for larger values of this ratio there is a window between the
two onsets. Kinks in CPE rate occur at these thresholds, marked by
dashed vertical lines in Fig. \ref{fig:comb}a for $E_C/\Delta=4$; in
between, $\gamma_{CPE}$ changes only little. One can see from this
plot that an optimum gate value - fast single-particle transfer and
errors only by CPE - exists for the case $E_C > \Delta$, and it lies
within $1/2+(2\Delta-eV)/4E_C<n_g<1-eV/4E_C$, closer to the upper
end of this range. The turnstile should thus be operated by a gate
voltage (Fig.~\ref{f0}b) switching between such an $n_g$ and
$1-n_g$.

\begin{figure}
\includegraphics[width=8.5cm]{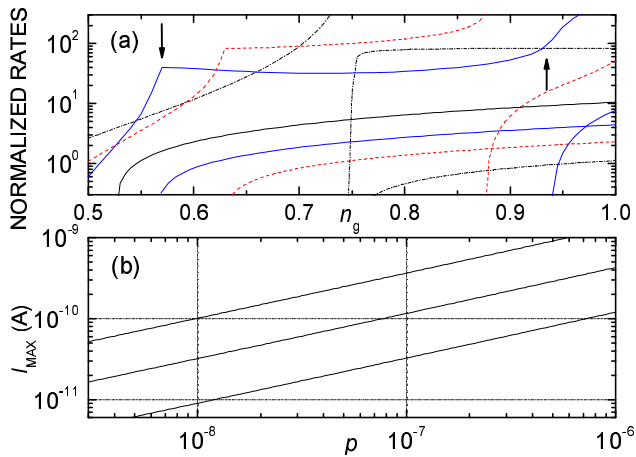}
\caption{ Performance of the SINIS hybrid turnstile. (a) Normalized
rates (for normalization, see text) of the various processes as
functions of gate position at the optimum bias point $eV =\Delta$:
CPE cotunneling (curves in the top part), single-particle (lower
right corner) and Andreev tunneling (taller curves on the right).
The different sets of curves refer to $E_C/\Delta = 1$ (black
dash-dotted), $2$ (red dashed), and $4$ (blue solid). The thresholds
at $E_C/\Delta = 4$ for single-particle and AR are indicated by
vertical arrows. (b) The maximum pumped current \protect (\ref{e7})
as a function of the allowed error rate $p$ for $E_C/\Delta=10, 4$
and $2$ from top to bottom.}
   \label{fig:comb}
\end{figure}

To make a quantitative estimate of the performance of the turnstile,
we note that since the CPE contributes one extra transferred
electron, the relative transfer error is $p=2\gamma_{CPE}/\gamma$ in
the operation window discussed above. This gives the (maximum)
junction conductance which can still suppress the CPE error to below
$p$ as $g =4\pi [ \mathcal{N} p \tilde{\gamma}/ \tilde{\gamma}_
{CPE}]^{1/2} $. On the other hand, one can drive the turnstile at a
frequency $f=\gamma/(2\ln(1/p))$ to suppress the missing cycle
errors to the same level. The maximum current of the turnstile at
the error rate $p$ is then
\begin{equation} \label{e7}
I_{\rm MAX}=ef =  \frac{e\Delta }{\hbar} \frac{ 2\pi}{\ln(1/p)} [
\mathcal{N} p \tilde{\gamma}^3/\tilde{\gamma}_{CPE}]^{1/2} .
\end{equation}
Figure \ref{fig:comb}b shows $I_{\rm MAX}$ versus $p$ for the most
common hybrid system using aluminium as the superconductor, for
which $\Delta/k_B \simeq 2.5$ K. In this plot, we also take into
account that $\mathcal{N}\propto E_C^{-1}$ because of the junction
area dependence of both of these quantities, and use an estimate of
the tunnel barrier characteristics $\mathcal{N} = 10^4$ for $E_C =
1$ K. We can see from Fig.~\ref{fig:comb} that increasing
$E_C/\Delta$ indeed improves the turnstile performance, and a single
turnstile with $E_C/\Delta =4$ reaches an accuracy of $10^{-8}$ at
about 30 pA current with $\simeq$ 400 k$\Omega$ junction resistance.
With $E_C/\Delta = 10$ (such high $E_C$:s were obtained, e.g., in
\cite{pashkin}), 100 pA current can be reached with the same
accuracy.

In summary, we have shown that a simple hybrid SINIS turnstile
should qualify as a metrological source of current. In order to
reach sufficient level of current, either a very large charging
energy or a few parallel turnstiles are needed. The latter option is
affordable because of the simplicity of the basic device \cite{b1}.
In practical pumps \cite{b6}, other sources of fluctuations (e.g.,
variations of the background charge) that can not be precisely
predicted by theory, influence the performance as well, although the
simplicity of our turnstile should make it stable also with respect
to these fluctuations.

This work was supported in part by NSF grant \# DMR-0325551, by
Technology Industries of Finland Centennial Foundation, and by the
Academy of Finland. We thank M. M\"ott\"onen for discussions.


\begin{thebibliography}{99}

\bibitem{al91} D.V. Averin and K.K. Likharev, in: {\sl Mesoscopic
phenomena in solids}, ed.\ by B.L. Altshuler, P.A. Lee and R.A.
Webb, (North-Holland, 1991), p.\ 173.

\bibitem{b2} L.J. Geerligs, V.F. Anderegg, P.A.M. Holweg, J.E.
Mooij, H. Pothier, D. Esteve, C. Urbina, and M.H. Devoret, Phys.\
Rev.\ Lett. {\bf 64}, 2691 (1990).

\bibitem{b3} H. Pothier, P. Lafarge, C. Urbina, D. Esteve,
and M.H. Devoret, Europhys.\ Lett. {\bf 17}, 249 (1992).

\bibitem{kel99} M.W. Keller, A.L. Eichenberger, J.M. Martinis, and
N.M. Zimmerman, Science {\bf 285}, 1706 (1999).

\bibitem{b10} S.V. Lotkhov, S.A. Bogoslovsky, A.B. Zorin, and J.
Niemeyer, Appl.\ Phys.\ Lett. {\bf 78}, 946 (2001).

\bibitem{sh96} J.M. Shilton {\sl et al.}, J. Phys.: Condens.
Matter {\bf 8}, L531 (1996).

\bibitem{fuj04} A. Fujiwara, N.M. Zimmerman, Y. Ono, and Y.
Takahashi, Appl.\ Phys.\ Lett. {\bf 84}, 1323 (2004).

\bibitem{bl07} M.D. Blumenthal {\sl et al.},
Nature Physics {\bf 3}, 343 (2007).

\bibitem{ka07} B. Kaestner {\sl et al.},
arXiv:0707.0993.

\bibitem{b1} J.P. Pekola, J.J. Vartiainen, M. M\"ott\"onen, O.-P.
Saira, M. Meschke, and D.V. Averin, Nature Physics {\bf 4}, 120
(2008).

\bibitem{b4} D.V. Averin, A.A. Odintsov, and S.V. Vyshenskii, J.\
Appl.\ Phys. {\bf 73}, 1297 (1993).

\bibitem{b5} D.V. Averin and Yu.V. Nazarov, Phys.\ Rev.\ Lett. {\bf
69}, 1993 (1992).

\bibitem{wil} J.W. Wilkins, in: {\sl Tunneling phenomena in solids},
ed. by E. Burnstein and S.Lundquist, (Plenum, 1969) p.\ 333.

\bibitem{cbt} Yu.V. Nazarov, J.\ Low Temp.\ Phys. {\bf 90}, 77
(1993); D.V. Averin, Physica B {\bf 194/196}, 979 (1994); H.
Schoeller and G. Sch\"{o}n, Phys.\ Rev. B {\bf 50}, 18436 (1994).

\bibitem{pashkin} Yu. A. Pashkin, Y. Nakamura, and J. S. Tsai,
Appl. Phys. Lett. {\bf 76}, 2256 (2000).

\bibitem{b6} R.L. Kautz, M.W. Keller, and J.M. Martinis, Phys.\
Rev.\ B {\bf 62}, 15888 (2000); X. Jehl, M.W. Keller, R.L. Kautz, J.
Aumentado, and J.M. Martinis, {\sl ibid.} {\bf 67}, 165331 (2003).

\end{thebibliography}
\end{document}